\documentclass[onecolumn]{elsart}    % Enable this line and disable the 
\usepackage{graphicx}          
\usepackage{amsmath}
\usepackage{amssymb} 
\begin{document}
\begin{frontmatter}

\title{On Euler Emulation of Observer-Based Stabilizers for Nonlinear Time-Delay Systems\thanksref{footnoteinfo}} 
                                                % than 10 words.

\thanks[footnoteinfo]{This work is supported in part by the Italian MIUR PRIN Project 2009, the Atheneum Project RIA 2016, and by the Center of Excellence for Research DEWS. Tel.: +39 0862434422; fax: +39 0862433180.}

\author[First]{M. Di Ferdinando}\ead{mario.diferdinando@graduate.univaq.it}\,\,\,\,\,\,\,\qquad\qquad  \qquad      % Add the 
\author[First]{P. Pepe}\ead{pierdomenico.pepe@univaq.it}               % e-mail address 

\address[First]{Department of Information Engineering, Computer Science, and Mathematics, University of L'Aquila, Via Vetoio (Coppito 1), 67100 L'Aquila, Italy.}


\begin{abstract}                          % Abstract of not more than 200 words.
In this paper, we deal with the problem of the stabilization in the sample-and-hold sense, by emulation of continuous-time, observer-based, global stabilizers. Fully nonlinear time-delay systems are studied. Sufficient conditions are provided such that the Euler approximation of continuous-time, observer-based, global stabilizers, for nonlinear time-delay systems, yields stabilization in the sample-and-hold sense. {\it Submitted (in an extended version) to Automatica.}
\end{abstract}

\end{frontmatter}
%===============================================================================
\section{Introduction} \label{introduction}

The emulation approach, for the implementation of controllers, is often the common choice in practical applications. In this approach, a continuous-time controller for the system at hand is firstly designed, ignoring sampling, and then, it is implemented digitally.  Sampled-data stabilization of linear, bilinear and nonlinear systems, even infinite dimensional ones, has been studied in the literature by many approaches,
 such as: i) the time-varying delay approach (see \cite{Fridman01}, \cite{Fridman02}  and \cite{Fridman03}), ii) the approximate system discretization approach (see \cite{Chen01}, \cite{Grune01}, \cite{Laila01}, \cite{NesicGrune01}, \cite{NesicTeel01}, \cite{NesicTeel02}, \cite{NesicTeel03}, \cite{Omran01}, \cite{Seuret01}, \cite{Briat03}, \cite{Seuret02}, \cite{Seuret03}), iii) the hybrid system approach (see \cite{Briat02} \cite{Carnevale01}, \cite{Hespanha01}, \cite{Naghshtabrizi01}, \cite{Naghshtabrizi02}, \cite{NesicTeel04}, \cite{NesicTeel05}, \cite{NesicTeel06});  iv) the stabilization in the sample-and-hold sense approach (see \cite{Clarke01}, \cite{Clarke02}, \cite{DiFerdinando02}, \cite{DiFerdinando01},  \cite{Pepe01}, \cite{Pepe02}, \cite{Pepe10}, \cite{Pepe03}). The reader can refer to \cite{Hetel01} for an interesting survey on the topic.
As far as nonlinear time-delay systems are concerned, state feedback sampled-data controllers are studied in  \cite{Karafyllis01}, \cite{Karafyllis02}, for nonlinear delay-free systems affected by time-delays in the input/output channels. More recently, the theory of  state feedback  stabilization in the sample-and-hold sense  (see \cite{Clarke01})  has been extended to fully nonlinear systems with state delays (see \cite{Pepe02},  \cite{Pepe03}).  As far as  any kind of stability preservation under sampling,  for fully nonlinear systems with state delays is concerned, to our best knowledge, no results are available in the literature for the case of emulated observer-based, continuous-time stabilizers. Actually, for fully nonlinear systems with state delays,  no theoretical stability results are available in the literature, for any kind of sampled-data observer-based controller.
In this paper, sufficient conditions are provided, such that continuous-time, observer-based, global stabilizers are stabilizers in the sample-and-hold sense. In particular, it is shown for fully nonlinear time-delay systems that, under suitable conditions, emulation, by Euler approximation, of nonlinear observer-based  global stabilizers, designed in the continuous time, yield stabilization in the sample-and-hold sense.

\medskip

{\bf Notation} $N$ denotes the set of nonnegative integer numbers, $\mathbb{R}$ denotes the set of real numbers, $\mathbb{R}^{\star}$ denotes the extended real line $\left[-\infty,+\infty\right]$, $\mathbb{R}^{+}$ denotes the set of nonnegative reals $\left[0,+\infty\right)$. The symbol $\left|\cdot\right|$ stands for the Euclidean norm of a real vector, or the induced Euclidean norm of a matrix. For a given positive integer $n$, for a symmetric, positive definite matrix $P\in \mathbb{R}^{n\times n}$, $\lambda_{max}\left(P\right)$ and $\lambda_{min}\left(P\right)$ denote the maximum and the minimum eigenvalue of $P$, respectively. For a given positive integer $n$ and a given positive real $h$, the symbol $\mathcal{B}_{h}^{n}$ denotes the subset $\{x \in \mathbb{R}^n:\ \vert x \vert \leq h  \}$. The essential supremum norm of an essentially bounded function is indicated with the symbol $\left\Vert \cdot\right\Vert _{\infty}$. For a positive integer $n$, for a positive real $\Delta$ (maximum involved time-delay): $\mathcal{C}^n$ denotes the space of the continuous functions mapping $\left[-\Delta,0\right]$ into $\mathbb{R}^{n}$; $W_{n}^{1,\infty}$ denotes the space of the absolutely continuous functions, with essentially bounded derivative, mapping $\left[-\Delta,0\right]$ into $\mathbb{R}^{n}$. For a positive real $p$, for $\phi\in\mathcal{C}^n$, $\mathcal{C}^{n}_{p}\left(\phi\right)=\left\{ \psi\in\mathcal{C}^n:\left\Vert \psi-\phi\right\Vert _{\infty}\leq p\right\} $. The symbol $\mathcal{C}^{n}_{p}$ denotes $\mathcal{C}^{n}_{p}\left(0\right)$. For a continuous function $x:\left[-\Delta,c\right)\rightarrow \mathbb{R}^{n}$, with $0<c\leq+\infty$, for any real $t\in\left[0,c\right)$, $x_{t}$ is the function in $\mathcal{C}^n$ defined as $x_{t}\left(\tau\right)=x\left(t+\tau\right),\ \tau\in\left[-\Delta,0\right]$.  For a positive integer $n$: $C^{1}\left(\mathbb{R}^{n};\mathbb{R}^{+}\right)$ denotes the space of the continuous functions from $\mathbb{R}^{n}$ to $\mathbb{R}^{+}$, admitting continuous (partial) derivatives; $C_{L}^{1}\left(\mathbb{R}^{n};\mathbb{R}^{+}\right)$ denotes the subset of the functions in $C^{1}\left(\mathbb{R}^{n};\mathbb{R}^{+}\right)$ admitting locally Lipschitz (partial) derivatives; $C^{1}\left(\mathbb{R}^{+};\mathbb{R}^{+}\right)$ denotes the space of the continuous functions from $\mathbb{R}^{+}\rightarrow \mathbb{R}^{+}$, admitting continuous derivative;  $C_{L}^{1}\left(\mathbb{R}^{+};\mathbb{R}^{+}\right)$ denotes the subset of functions in $C^{1}\left(\mathbb{R}^{+};\mathbb{R}^{+}\right)$ admitting locally Lipschitz derivative. Let us here recall that a continuous function $\gamma:\mathbb{R}^{+}\rightarrow \mathbb{R}^{+}$ is: of class $\mathcal{P}_{0}$ if $\gamma\left(0\right)=0$; of class $\mathcal{P}$ if it is of class $\mathcal{P}_{0}$ and $\gamma\left(s\right)>0$, $s>0$; of class $\mathcal{K}$ if it is of class $\mathcal{P}$ and strictly increasing; of class $\mathcal{K}_{\infty}$ if it is of class $\mathcal{K}$ and unbounded; of class $\mathcal{L}$ if  it monotonically decreases to zero as its argument tends to $+\infty$. A continuous function $\beta:\mathbb{R}^{+}\times \mathbb{R}^{+}\rightarrow \mathbb{R}^{+}$ is of class $\mathcal{KL}$ if, for each fixed $t\geq0$, the function $s\to\beta\left(s,t\right)$ is of class $\mathcal{K}$  and, for
each fixed $s\geq0$, the function $t\to\beta\left(s,t\right)$ is of class $\mathcal{L}$. For positive integers $n$, $m$, for a map $f:\mathcal{C}^n\times \mathbb{R}^{m}\rightarrow \mathbb{R}^{n}$, and  for a locally Lipschitz functional $V:\mathcal{C}^n\rightarrow \mathbb{R}^{+}$, the derivative in Driver's form (see \cite{Pepe05} and the references therein) $D^{+}V:\mathcal{C}^n\times \mathbb{R}^{m}\rightarrow \mathbb{R}^{\star}$, of the functional $V$, is defined, for $\phi\in\mathcal{C}^n$, $u\in \mathbb{R}^{m}$, as:
\begin{equation} \label{derivata01} 
D^{+}V\left(\phi,u\right)=\limsup_{h\rightarrow0^{+}}\frac{V\left(\phi_{h,u}\right)-V\left(\phi\right)}{h},
\end{equation}
where, for $0\leq h<\Delta$, $\phi_{h,u}\in\mathcal{C}^n$ is defined, for $s \in \left[-\Delta,0\right]$, as \\
\begin{equation} \label{derivata03} 
\phi_{h,u}\left(s\right)=\begin{cases}
\phi\left(s+h\right), & s\in\left[-\Delta,-h\right),\\
\phi\left(0\right)+\left(s+h\right)f\left(\phi,u\right), & s\in\left[-h,0\right]. \nonumber
\end{cases}
\end{equation} \\ 
Throughout the paper, GAS stands for globally asymptotically stable or global asymptotic stability, RFDE stands for retarded functional differential equation, CLKF stands for control Lyapunov-Krasovskii functional.
\section{Emulation of Observer-Based Stabilizers for fully Nonlinear Time-Delay Systems}
In this section, we present our main result concerning sampled-data stabilization  for nonlinear time-delay systems, stabilized by continuous-time observer-based controllers. In particular, taking into account the stabilization in the sample-and-hold sense theory (see \cite{DiFerdinando01}, \cite{Pepe02}, \cite{Pepe10}, \cite{Pepe03}), it is shown that, under suitable conditions: there exists a minimal sampling frequency (aperiodic sampling is allowed) such that, by Euler emulation of an observer-based controller, semi-global practical stability, with arbitrary small final target ball of the origin, is guaranteed.
\subsection{Plant and continuous-time controller description}
Let us consider a fully nonlinear time-delay system, described by the following RFDE (see \cite{Hale01}, \cite{Kolmanovskii01})
\begin{equation} \label{modellodelay01} 
\begin{array}{ll}
\overset{.}{x}\left(t\right)=f\left(x_t,u\left(t\right)\right),\,\,\,\,t\geq0\,\,\,\,a.e.,
& \\ y\left(t\right)=h\left(x_t\right),
& \\ x\left(\tau\right)=x_0\left(\tau\right),\,\,\,\,\tau\in\left[-\Delta,0\right],
\end{array}
\end{equation}
where: $x\left(t\right)\in \mathbb{R}^n$, $n$ is a positive integer; $x_0$, $x_t\in\mathcal{C}^n$; $\Delta$ is a positive real, the maximum involved time delay; $u\left(t\right)\in \mathbb{R}^{m}$ is the input (the input signal is Lebesgue measurable and locally essentially bounded); $m$ is a positive integer; $y\left(t\right)\in \mathbb{R}^{q}$ is the output, $q$ is a positive integer; $f$ is a map from $\mathcal{C}^n\times \mathbb{R}^{m}$ to $\mathbb{R}^{n}$, Lipschitz on bounded sets; $h$ is a map from $\mathcal{C}^n$ to $\mathbb{R}^{q}$, Lipschitz on bounded sets. It is assumed that $f\left(0,0\right)=h(0)=0$. Furthermore, it is assumed that the initial state $x_{0}\in W_{n}^{1,\infty}$, and that $ess\sup_{\theta\in\left[-\Delta,0\right]}\left|\frac{dx_{0}\left(\theta\right)}{d\theta}\right|\leq \frac{\tilde{q}}{\sqrt{2}}$ (see \cite{Pepe02}, \cite{Pepe03}), where $\tilde{q}$ is a given, arbitrary, positive real (the utility of the term $\sqrt{2}$ will be clear from forthcoming Remark \ref{assumdelay03}). We recall that the system described by (\ref{modellodelay01}) admits a locally absolutely continuous solution in a maximal time interval $\left[0,b\right)$, with $0<b\leq+\infty$ (see \cite{Hale01}). \\
Let us  consider now an observer-based controller  for the nonlinear time-delay system  (\ref{modellodelay01}), described by the following equations (see \cite{Germani04}, \cite{Germani05}, \cite{Germani02}, \cite{Zhang01})
\begin{equation} \label{modellodelay02} 
\begin{array}{ll}
\overset{.}{\hat{x}}\left(t\right)=\widehat{f}\left(\hat{x}_t,u\left(t\right),y\left(t\right)\right),\,\,\,\,t\geq0,
& \\ u\left(t\right)=k\left(\hat{x}_t,y(t)\right),
& \\ \hat{x}\left(\tau\right)=\hat{x}_0\left(\tau\right),\,\,\,\,\tau\in\left[-\Delta,0\right],
\end{array}
\end{equation}
where: $\hat{x}\left(t\right)\in \mathbb{R}^n$; $\hat{x}_0$, $\hat{x}_t\in\mathcal{C}^n$; $\Delta$ is the maximum involved time delay as in (\ref{modellodelay01}); $u\left(t\right)\in \mathbb{R}^{m}$ and $y\left(t\right)\in \mathbb{R}^{q}$ are the input and the output as defined in (\ref{modellodelay01}), respectively; the maps  $\widehat{f}:\mathcal{C}^n\times \mathbb{R}^{m}\times \mathbb{R}^{q}\rightarrow \mathbb{R}^{n}$ and $k:\mathcal{C}^n\times \mathbb{R}^{q}\rightarrow \mathbb{R}^{m}$ are Lipschitz on bounded sets; it is assumed that $\widehat{f}\left(0,0,0\right)=k\left(0,0\right)=0$. Taking into account the assumption on the initial state $x_0$, it is assumed that the initial state $\hat{x}_{0}\in W_{n}^{1,\infty}$, and that $ess\sup_{\theta\in\left[-\Delta,0\right]}\left|\frac{d\hat{x}_{0}\left(\theta\right)}{d\theta}\right|\leq \frac{\tilde{q}}{\sqrt{2}}$ (see \cite{Pepe02}, \cite{Pepe03}).  From (\ref{modellodelay01}), (\ref{modellodelay02})  it readily follows that the related closed-loop system is described by the RFDE
\begin{equation} \label{modellodelayclosedloop001} 
\begin{array}{ll}
\overset{.}{x}\left(t\right)=f\left(x_t,k\left(\hat{x}_t,h\left(x_t\right)\right)\right),\,\,\,\,t\geq0,
& \\ \overset{.}{\hat{x}}\left(t\right)=\widehat{f}\left(\hat{x}_t,k\left(\hat{x}_t,h\left(x_t\right)\right),h\left(x_t\right)\right),
& \\ x\left(\tau\right)=x_0\left(\tau\right),\,\,\,\,\hat{x}\left(\tau\right)=\hat{x}_0\left(\tau\right),\,\,\,\,\tau\in\left[-\Delta,0\right].
\end{array}
\end{equation}
Let (as long as the solution of (\ref{modellodelayclosedloop001}) exists)
\begin{equation} \label{modellodelayx04} 
\begin{array}{ll}
\tilde{x}\left(t\right)=\begin{bmatrix}x\left(t\right)\\
\hat{x}\left(t\right)
\end{bmatrix}\in \mathbb{R}^{2n},\,\,\,\,\,\,\,\,\tilde{x}_t=\begin{bmatrix}x_t\\
\hat{x}_t
\end{bmatrix}\in \mathcal{C}^{2n}.
\end{array}
\end{equation}
From the closed-loop system (\ref{modellodelayclosedloop001}), taking into account (\ref{modellodelayx04}), we have that
\begin{equation} \label{modellodelayaggiunta1234} 
\begin{array}{ll}
\overset{.}{\tilde{x}}\left(t\right)=F(\tilde{x}_t),
\end{array}
\end{equation}
where $F:\mathcal{C}^{2n}\rightarrow \mathbb{R}^{2n}$ is defined, for $\tilde{\phi}=\begin{bmatrix}\tilde{\phi}_1 \\ \tilde{\phi}_2\end{bmatrix}\in \mathcal{C}^{2n}$, $\tilde{\phi}_i\in\mathcal{C}^{n}$, $i=1,2$, as
\begin{equation} \label{modellodelayaggiunta12345}
\begin{array}{ll}
F(\tilde{\phi})=\begin{bmatrix}f(\tilde{\phi}_1,k(\tilde{\phi}_2,h(\tilde{\phi}_1)))\\
\widehat{f}(\tilde{\phi}_2,k(\tilde{\phi}_2,h(\tilde{\phi}_1)),h(\tilde{\phi}_1))
\end{bmatrix}.
\end{array}
\end{equation}
\begin{rem} \label{assumdelay03}
Notice that, taking into account (\ref{modellodelayx04}) and the assumption on the initial states $x_0$ and $\hat{x}_0$, the initial state $\tilde{x}_0\in W_{2n}^{1,\infty}$, and, moreover $$ess\sup_{\theta\in\left[-\Delta,0\right]}\left|\frac{d\tilde{x}_0\left(\theta\right)}{d\theta}\right|\leq \tilde{q}$$.
\end{rem}
%%%%%%%%%%%%%%%%%%%%%%%%%%%%%%%%%%%%%%%%%%%%%%%%%%%%%%%%%%%%%%%%%%%%%%%%%%%%%%%%
\subsection{Emulation of observer-based stabilizers}
Firstly, we recall the definition of smoothly separable functionals,  which are helpful in sampled-data stabilization of time-delay systems (see \cite{Pepe02}, \cite{Pepe03}). 
\begin{defn} \label{smoothlyseparable} 
(see \cite{Pepe02}, \cite{Pepe03}) A functional $V:\mathcal{C}^n\rightarrow \mathbb{R}^{+}$ is said to be smoothly separable if there exist a function $V_{1}\in C_{L}^{1}\left(\mathbb{R}^{n};\mathbb{R}^{+}\right)$, a locally Lipschitz functional $V_{2}:\mathcal{C}^n\rightarrow\mathbb{R}^{+}$, functions $\beta_{i}$ of class $\mathcal{K_{\infty}}$, $i = 1, 2$, such that, for any $\phi\in\mathcal{C}^n$, the following equality/inequalities hold
\begin{equation} \label{smoothlyseparable01} 
\begin{array}{ll}
V\left(\phi\right)=V_{1}\left(\phi\left(0\right)\right)+V_{2}\left(\phi\right),
& \\ \beta_{1}\left(\left|\phi\left(0\right)\right|\right)\leq V_{1}\left(\phi\left(0\right)\right)\leq\beta_{2}\left(\left|\phi\left(0\right)\right|\right).
\end{array}
\end{equation}
\end{defn}
Let us introduce the following assumption for the closed-loop system described by (\ref{modellodelayaggiunta1234}).
\begin{assum} \label{assumdelay02}
(see \cite{Pepe02}, \cite{Pepe03}) There exist a smoothly separable functional $V:\mathcal{C}^{2n}\rightarrow\mathbb{R}^{+}$, functions $\gamma_{1}$, $\gamma_{2}$ of class $\mathcal{K_{\infty}}$, positive reals $\eta$, $\mu$, a function $p$ in $C_{L}^{1}\left(\mathbb{R}^{+};\mathbb{R}^{+}\right)$, of class $\mathcal{K_{\infty}}$, $\nu\in\left\{ 0,1\right\}$, a function $\alpha_3$ of class $\mathcal{K}$, such that:
\begin{itemize}
\item[1)] the following inequalities (with respect to the system described by (\ref{modellodelayaggiunta1234})) hold, for any $\tilde{\phi}\in\mathcal{C}^{2n}$,
\begin{equation} \label{steepestdescentfeedback03} 
\begin{array}{ll}
\gamma_{1}(|\tilde{\phi}(0)|)\leq V(\tilde{\phi})\leq\gamma_{2}(\Vert \tilde{\phi}\Vert _{\infty}),
 & \\ D^{+}V(\tilde{\phi},0)\leq-\alpha_3(|\tilde{\phi}(0)|);
\end{array}
\end{equation}
\item[2)] the following inequality (with respect to the system described by (\ref{modellodelayaggiunta1234})) holds, for any $\tilde{\phi}\in\mathcal{C}^{2n}$,
\begin{equation} \label{steepestdescentfeedback04}
\begin{array}{ll}
\nu D^{+}V(\tilde{\phi},0)+\eta D^{+}p\circ V_{1}(\tilde{\phi},0)+\eta\mu p\circ V_{1}(\tilde{\phi}\left(0\right))\leq0.
\end{array}
\end{equation}
\end{itemize}
The map $(\tilde{\phi},\tilde{u})\rightarrow D^{+}V_2(\tilde{\phi},\tilde{u})$, $\tilde{\phi}\in\mathcal{C}^{2n}$, $\tilde{u}\in \mathbb{R}^{m+n}$,  is Lipschitz on bounded subsets of $\mathcal{C}^{2n}\times \mathbb{R}^{m+n}$.
\end{assum}
Notice that, in (\ref{steepestdescentfeedback04}),  the derivative in Driver's form of the functional $V$, with respect to the input-free system described by (\ref{modellodelayaggiunta1234}), is used. Thus, taking into account (\ref{derivata01}),  the fictitious choice  $u=0$ is taken for the derivative in Driver's form of $V$ in (\ref{steepestdescentfeedback04}). The same choice is taken for the term $D^+ p\circ V_1(\tilde{\phi},0)$.
The main result of the paper, for fully nonlinear time-delay systems, is provided by the following theorem.
\begin{thm}\label{mainresult02}
Let Assumption \ref{assumdelay02} hold. Let $a$ be an arbitrary real in $\left(0,1\right]$. Then, for any positive reals $R$, $r$ with $0<r<R$, there exist positive reals $\delta$, $T$, $E$ such that, for any partition $\pi_{a,\delta}=\left\{ t_{j},\,\, j=0,1,...\right\} $, for any initial states $x_0$, $\hat{x}_0$ such that $\tilde{x}_0=\begin{bmatrix}x_0\\
\hat{x}_0
\end{bmatrix}\in\mathcal{C}^{2n}_{R}$, the corresponding solution of the sampled-data closed-loop system, described by the equations
\begin{equation} \label{closedloopsampleddelay001} 
\begin{array}{ll}
\overset{.}{x}\left(t\right)=f(x_t,k(\hat{x}_{t_{j}},h(x_{t_{j}}))),
& \\  \hat{x}\left(t_{j+1}\right)=\hat{x}\left(t_{j}\right)+\left(t_{j+1}-t_j\right)\widehat{f}(\hat{x}_{t_{j}},k(\hat{x}_{t_{j}},h(x_{t_{j}})),h(x_{t_{j}})),
& \\
& \\ \hat{x}_{t_{j}}(\theta)=
\begin{cases}
\hat{x}_0(t_j+\theta), & t_j+\theta\leq0,\\ \\
\hat{x}(t_{k})+\dfrac{t_j+\theta-t_k}{t_{k+1}-t_k}(\hat{x}(t_{k+1})-\hat{x}(t_{k})), & t_j+\theta>0, \\
t_k=\underset{l\in N}{\max}\{t_l\in\pi_{a,\delta}\,:\,t_l\leq t_j+\theta\},
\end{cases}
& \\ 
& \\ \theta\in[-\Delta, 0],\,\,\,\,t\in\left[t_j,t_{j+1}\right),\,\,\,\,j=0,1,...,
\end{array}
\end{equation}
exists for all $t\in \mathbb{R}^+$, $t_j\in\pi_{a,\delta}$, and, furthermore, satisfies:
\begin{equation} \label{result001} 
\begin{array}{ll}
\begin{bmatrix}x_t\\
\hat{x}_{t_{j}}
\end{bmatrix}\in \mathcal{C}^{2n}_{E},\,\,\,\,\forall t\in \mathbb{R}^+,\,\,\,\,\forall t_j\in\pi_{a,\delta},\\
& \\ \begin{bmatrix}x_t\\
\hat{x}_{t_{j}}
\end{bmatrix}\in \mathcal{C}^{2n}_{r},\,\,\,\,\forall t\geq T,\,\,\,\,\forall t_j\in\pi_{a,\delta},\,\,\,\, t_j\geq T.
\end{array}
\end{equation}
\end{thm}
\begin{pf}
The proof is reported in Section \ref{proof02} of the Appendix.
\end{pf}
%%%%%%%%%%%%%%%%%%%%%%%%%%%%%%%%%%%%%%%%%%%%%%%%%%%%%%%%%%%%%%%%%%%%%%%%%%%%%%%%%%%%%%%%%%%%%%%%%%%%%%%%%%%%%%%%%%%%
\section{Conclusions}
In this paper, it has been shown for fully nonlinear time-delay systems that, under suitable conditions, emulation, by Euler approximation, of nonlinear observer-based global stabilizers designed in the continuous time, guarantees stabilization in the sample-and-hold sense.

%\bibliographystyle{plain}   
%\bibliography{autosam}
\appendix
\section{Proof of Theorem \ref{mainresult02}} \label{proof02} 
Firstly, for the reader convenience, we recall a class of Lyapunov-Krasovskii functionals, which are very helpful for sampled-data stabilization. In particular, we report the definition concerning CLKFs (see \cite{Jankovic01}, \cite{Karafyllis03}, \cite{Pepe02}, \cite{Pepe14}, \cite{Pepe15}, \cite{Sontag01}). Furthermore, we report the definition of  steepest descent feedback induced by a CLKF (see \cite{Clarke02}, \cite{Pepe02}, \cite{Pepe03}), and of stabilizers in the sample-and-hold sense (see \cite{Clarke01}, \cite{Clarke02}, \cite{Pepe02}, \cite{Pepe03}).
\begin{defn}  \label{CLKFdef} 
(see \cite{Jankovic01}, \cite{Karafyllis03}, \cite{Pepe02}, \cite{Pepe14}, \cite{Pepe15}, \cite{Sontag01}) A smoothly separable functional $V:\mathcal{C}^n\rightarrow \mathbb{R}^{+}$ is said to be a CLKF, for the system described by (\ref{modellodelay01}), if there exist functions $\gamma_{1}$, $\gamma_{2}$ of class $\mathcal{K_{\infty}}$ such that the following inequalities hold
\begin{equation} \label{CLKF} 
\begin{array}{ll}
\gamma_{1}\left(\left|\phi\left(0\right)\right|\right)\leq V\left(\phi\right)\leq\gamma_{2}\left(\left\Vert \phi\right\Vert _{\infty}\right),\,\,\,\,\forall\phi\in\mathcal{C}^n,
& \\ \inf_{u\in \mathbb{R}^m}D^{+}V\left(\phi,u\right)<0,\,\,\,\,\forall\phi\in\mathcal{C}^n,\,\,\,\,\phi\left(0\right)\neq0.
\end{array}
\end{equation}
\end{defn}
\begin{defn}\label{steepestdescentfeedbackk} 
(see \cite{Clarke02}, \cite{Pepe02}, \cite{Pepe03}) A map $k:\mathcal{C}^n\rightarrow \mathbb{R}^m$ (continuous or not) is said to be a steepest descent feedback, induced by a CLKF $V$, for the system described by (\ref{modellodelay01}), if the following condition holds:
 there exist positive reals $\eta$, $\mu$, a function $p$ in $C_{L}^{1}\left(\mathbb{R}^{+};\mathbb{R}^{+}\right)$, of class $\mathcal{K_{\infty}}$, $\nu\in\left\{ 0,1\right\}$, a function $\bar{\alpha}$ of class $\mathcal{K}$ such that $I_{d}-\bar{\alpha}$ is of class $\mathcal{K_{\infty}}$, such that the inequality holds
\begin{equation} \label{steepestdescentfeedback01} 
\begin{array}{ll}
\nu D^{+}V\left(\phi,k\left(\phi\right)\right)+\eta\max\left\{ 0,D^{+}p\circ V_{1}\left(\phi,k\left(\phi\right)\right)+\mu p\circ V_{1}\left(\phi\left(0\right)\right)\right\}
& \\ \leq\bar{\alpha}\left(\eta\mu e^{-\mu\Delta}p\circ\beta_{1}\left(\left\Vert \phi\right\Vert _{\infty}\right)\right),
\end{array}
\end{equation}
with $\beta_1$ given in Definition \ref{smoothlyseparable}.
\end{defn}
\begin{defn} 
(see \cite{Clarke01}, \cite{Clarke02}, \cite{Pepe02}, \cite{Pepe03}) We say that a feedback $K:\mathcal{C}^n\rightarrow \mathbb{R}^m$ (continuous or not) stabilizes the system described by (\ref{modellodelay01}) in the sample-and-hold sense if, for every positive reals $r,R$, $0<r<R$, $a\in\left(0,1\right]$, there exist a positive real $\delta$ depending upon $r$, $R$, $\tilde{q}$ and $\Delta$, a positive real $T$, depending upon $r$, $R$, $\tilde{q}$, $\Delta$ and $a$, and a positive real $E$ depending upon $R$ and $\Delta$, such that, for any partition $\pi_{a,\delta}=\left \{t_i, i=0,1,...\right \}$, for any initial value $x_0\in \mathcal{C}^{n}_{R}$, the solution corresponding to $x_0$ and to the sampled-data feedback control law $u\left(t\right)=K(x_{t_j})$, $t_j\leq t < t_{j+1}$, $j=0,1,...,$ exists for all $t\geq0$ and, furthermore, satisfies
\begin{equation} 
x_t\in\mathcal{C}^{n}_{E}\,\,\,\,\forall t\geq0,\,\,\,\,x_t\in\mathcal{C}^{n}_{r}\,\,\,\,\forall t\geq T.\nonumber
\end{equation}
\end{defn}
In the following, we recall the main theorem in \cite{Pepe02}. 
\begin{thm}  \label{thmtimedelaySAH}
(see Theorem 5.3 in \cite{Pepe02}) Any steepest descent feedback $k$ (continuous or not, see Definition \ref{steepestdescentfeedbackk}) stabilizes the system described by (\ref{modellodelay01}) in the sample-and-hold sense.
\end{thm}
Taking into account (\ref{modellodelay01}) and (\ref{modellodelay02}), let us consider the open-loop system described by the following RFDEs
\begin{equation} \label{modellodelay03} 
\begin{array}{ll}
\overset{.}{x}\left(t\right)=f\left(x_t,\tilde{u}_1\left(t\right)\right),
& \\ \overset{.}{\hat{x}}\left(t\right)=\tilde{u}_2\left(t\right),\,\,\,\,t\geq0\,\,\,\,a.e.,
& \\ y\left(t\right)=h\left(x_t\right),
& \\ x\left(\tau\right)=x_0\left(\tau\right),\,\,\,\,\hat{x}\left(\tau\right)=\hat{x}_0\left(\tau\right),\,\,\,\,\tau\in\left[-\Delta,0\right],
\end{array}
\end{equation}
where: $x_0$, $\hat{x}_0\in\mathcal{C}^{n}$ are the initial states in (\ref{modellodelay01}) and (\ref{modellodelay02}); $x_t$, $\hat{x}_t\in\mathcal{C}^{n}$; $x\left(t\right),\hat{x}\left(t\right)\in \mathbb{R}^n$; $f$ is the map in (\ref{modellodelay01}); $\tilde{u}_1\left(t\right)=u\left(t\right) \in \mathbb{R}^m$ is the input in (\ref{modellodelay01}); $\tilde{u}_2\left(t\right) \in \mathbb{R}^n$ is a new input. The input signals are Lebesgue measurable and locally essentially bounded; $y\left(t\right)\in \mathbb{R}^q$ is the output in (\ref{modellodelay01}); $h$ is the map in (\ref{modellodelay01}).  Taking into account (\ref{modellodelayx04}), let (as long as the solution of (\ref{modellodelay03}) exists)
\begin{equation} \label{modellodelay04} 
\begin{array}{ll}
\tilde{x}\left(t\right)=\begin{bmatrix}x\left(t\right)\\
\hat{x}\left(t\right)
\end{bmatrix}\in \mathbb{R}^{2n},\,\,\,\, \tilde{x}_t=\begin{bmatrix}x_t\\
\hat{x}_t
\end{bmatrix}\in \mathcal{C}^{2n},\,\,\,\,\tilde{u}\left(t\right)=\begin{bmatrix}\tilde{u}_1\left(t\right)\\
\tilde{u}_2\left(t\right)
\end{bmatrix}\in \mathbb{R}^{m+n}.
\end{array}
\end{equation}
By (\ref{modellodelay04}), system (\ref{modellodelay03}) can be written as follows
\begin{equation} \label{modellodelay05} 
\begin{array}{ll}
\overset{.}{\tilde{x}}\left(t\right)=\begin{bmatrix}\overset{.}{x}\left(t\right)\\
\overset{.}{\hat{x}}\left(t\right)
\end{bmatrix}=\begin{bmatrix}f\left(x_t,\tilde{u}_1\left(t\right)\right)\\
\tilde{u}_2\left(t\right)
\end{bmatrix}=\widetilde{F}\left(\tilde{x}_t,\tilde{u}\left(t\right)\right),
& \\ \tilde{x}\left(\tau\right)=\tilde{x}_0\left(\tau\right)=\begin{bmatrix}x_0\left(\tau\right)\\
\hat{x}_0\left(\tau\right)
\end{bmatrix}\,\,\,\,\tau\in\left[-\Delta,0\right],
\end{array}
\end{equation}
where the map  $\widetilde{F}:\mathcal{C}^{2n}\times \mathbb{R}^{m+n}\rightarrow \mathbb{R}^{2n}$ is readily defined by (\ref{modellodelay05}). The map $\widetilde{F}$ is Lipschitz on bounded sets and $\widetilde{F}\left(0,0\right)=0$. Taking into account the observer-based controller (\ref{modellodelay02}), let
\begin{equation} \label{feedbackdelay001} 
\begin{bmatrix}k\left(\hat{x}_t,h\left(x_t\right)\right)\\
\widehat{f}\left(\hat{x}_t,k\left(\hat{x}_t,h\left(x_t\right)\right),h\left(x_t\right)\right)
\end{bmatrix}=\tilde{k}\left(\tilde{x}_t\right),
\end{equation}
where: $\widehat{f}$ and $k$ are the maps in (\ref{modellodelay02}); $h$ is the map in (\ref{modellodelay01}); $\tilde{k}:\mathcal{C}^{2n}\rightarrow \mathbb{R}^{m+n}$ is readily defined by (\ref{feedbackdelay001}), is Lipschitz on bounded sets and satisfies $\tilde{k}\left(0\right)=0$. Taking into account (\ref{modellodelayaggiunta12345}), (\ref{modellodelay04}), (\ref{modellodelay05}), (\ref{feedbackdelay001}),   for any $\tilde{\phi}\in\mathcal{C}^{2n}$, the following equality holds
\begin{equation} \label{modellodelay05ok02}
\begin{array}{ll}
F(\tilde{\phi})=\widetilde{F}(\tilde{\phi},\tilde{k}(\tilde{\phi})).
\end{array}
\end{equation}
Therefore, the continuous-time closed-loop system (\ref{modellodelay01}), (\ref{modellodelay02}) is also described, besides by (\ref{modellodelayaggiunta1234}),  by the following RFDEs
\begin{equation} \label{modellodelay05ok01}
\begin{array}{ll}
\overset{.}{\tilde{x}}\left(t\right)=\widetilde{F}(\tilde{x}_t,\tilde{k}(\tilde{x}_t)),
& \\ \tilde{x}\left(\tau\right)=\tilde{x}_0\left(\tau\right)=\begin{bmatrix}x_0\left(\tau\right)\\
\hat{x}_0\left(\tau\right)
\end{bmatrix},\,\,\,\,\tau\in\left[-\Delta,0\right].
\end{array}
\end{equation}
Taking into account (\ref{steepestdescentfeedback03}), (\ref{steepestdescentfeedback04}) in Assumption \ref{assumdelay02}, for any $\tilde{\phi}\in\mathcal{C}^{2n}$, the following inequality (with respect to the system described by (\ref{modellodelayaggiunta1234})) holds
\begin{equation} \label{steepestdescentfeedback04kaggiunta}
\begin{array}{ll}
\nu D^{+}V(\tilde{\phi},0)+\eta\max\{ 0,D^{+}p\circ V_{1}(\tilde{\phi},0)+\mu p\circ V_{1}(\tilde{\phi}\left(0\right))\}\leq0.
\end{array}
\end{equation} \\
Taking into account (\ref{modellodelayaggiunta1234}), (\ref{steepestdescentfeedback04}), (\ref{modellodelay05}) (\ref{modellodelay05ok02}), (\ref{modellodelay05ok01}), (\ref{steepestdescentfeedback04kaggiunta}), we have, $\forall\tilde{\phi}\in\mathcal{C}^{2n}$ (with respect to the system described by (\ref{modellodelay05}))
\begin{equation} \label{steepestdescentfeedback04k}
\begin{array}{ll}
\nu D^{+}V(\tilde{\phi},\tilde{k}(\tilde{\phi}))+\eta\max\{ 0,D^{+}p\circ V_{1}(\tilde{\phi},\tilde{k}(\tilde{\phi}))+\mu p\circ V_{1}(\tilde{\phi}\left(0\right))\}\leq0.
\end{array}
\end{equation} \\
Indeed, the derivative in Driver's form of the functional $V$ (see (\ref{derivata01})), with respect to the system described by (\ref{modellodelayaggiunta1234}), is equal to the one of the functional $V$ with respect to the system described by (\ref{modellodelay05}), with $\tilde{u}(t)=\tilde{k}(\tilde{x}_t)$ (see (\ref{modellodelayaggiunta1234}), (\ref{modellodelay05}), (\ref{modellodelay05ok02}), (\ref{modellodelay05ok01})).
Then, taking again into account this fact, from (\ref{steepestdescentfeedback03}) it follows that the functional $V$ is a CLKF, for the system described by (\ref{modellodelay05}), according to Definition \ref{CLKFdef} (just consider the case $\tilde{u}=\tilde{k}(\tilde{\phi})$, $\tilde{\phi}\in\mathcal{C}^{2n}$). From (\ref{steepestdescentfeedback04k}), it follows that the state feedback $\tilde{k}$, in (\ref{feedbackdelay001}), is a steepest descent feedback induced by the CLKF $V$, for the system described by (\ref{modellodelay05}), according to Definition \ref{steepestdescentfeedbackk}. Then, by Theorem \ref{thmtimedelaySAH}, for any positive reals $R$, $r$ with $0<r<R$, there exist positive reals $\delta$, $T$, $E$ such that, for any partition $\pi_{a,\delta}=\left\{ t_{j},\,\, j=0,1,...\right\} $, for any initial states $x_0$, $\hat{x}_0$ such that $\tilde{x}_0=\begin{bmatrix}x_0\\
\hat{x}_0
\end{bmatrix}\in\mathcal{C}^{2n}_{R}$, the solution $\tilde{x}_t=\begin{bmatrix}x_t\\
\hat{x}_{t}
\end{bmatrix}$ of system (\ref{modellodelay05}), with 
\begin{equation} \label{feedbacksampleddelay}
\begin{array}{ll} 
\tilde{u}\left(t\right)=\tilde{k}(\tilde{x}_{t_{j}})=\begin{bmatrix}k(\hat{x}_{t_{j}},h(x_{t_{j}}))\\
\widehat{f}(\hat{x}_{t_{j}},k(\hat{x}_{t_{j}},h(x_{t_{j}})),h(x_{t_{j}}))
\end{bmatrix},
& \\ t_j\leq t<t_{j+1},\,\,\,\,t_j\in\pi_{a,\delta},\,\,\,\,j=0,1,...,
\end{array}
\end{equation}
exists for all $t\in \mathbb{R}^+$ and, furthermore, satisfies
\begin{equation} \label{result0012ok} 
\begin{array}{ll}
\tilde{x}_t=\begin{bmatrix}x_t\\
\hat{x}_{t}
\end{bmatrix}\in \mathcal{C}^{2n}_{E},\,\,\,\,\forall t\in \mathbb{R}^+,\,\,\,\,\tilde{x}_t=\begin{bmatrix}x_t\\
\hat{x}_{t}
\end{bmatrix}\in \mathcal{C}^{2n}_{r},\,\,\,\,\forall t\geq T.
\end{array}
\end{equation}
Now, from (\ref{modellodelay05}), (\ref{feedbackdelay001}) and (\ref{feedbacksampleddelay}), it follows that $\tilde{x}_t=\begin{bmatrix}x_t\\
\hat{x}_{t}
\end{bmatrix}$ is the solution, for $t\in \mathbb{R}^+$, of the closed-loop system described by the equations
\begin{equation} \label{closedloopsampleddelay} 
\begin{array}{ll}
\overset{.}{x}\left(t\right)=f(x_t,k(\hat{x}_{t_{j}},h(x_{t_{j}}))),
& \\ \overset{.}{\hat{x}}\left(t\right)=\widehat{f}(\hat{x}_{t_{j}},k(\hat{x}_{t_{j}},h(x_{t_{j}})),h(x_{t_{j}})),
& \\
& \\ \hat{x}_{t_{j}}(\theta)=
\begin{cases}
\hat{x}_0(t_j+\theta), & t_j+\theta\leq0,\\ \\
\hat{x}(t_{k})+\dfrac{t_j+\theta-t_k}{t_{k+1}-t_k}(\hat{x}(t_{k+1})-\hat{x}(t_{k})), & t_j+\theta>0, \\
t_k=\underset{l\in N}{\max}\{t_l\in\pi_{a,\delta}\,:\,t_l\leq t_j+\theta\},
\end{cases}
& \\
& \\ \theta\in[-\Delta, 0],\,\,\,\,t\in\left[t_j,t_{j+1}\right),\,\,\,\,t_j\in\pi_{a,\delta},\,\,\,\,j=0,1,...,
& \\ x\left(\tau\right)=x_0\left(\tau\right),\,\,\,\,\hat{x}\left(\tau\right)=\hat{x}_0\left(\tau\right),\,\,\,\,\tau\in\left[-\Delta,0\right].
\end{array}
\end{equation}
From (\ref{closedloopsampleddelay}), it follows that $\begin{bmatrix}x_t\\
\hat{x}_{t_{j}}
\end{bmatrix}$ is the solution, for $t\in \mathbb{R}^+$, $t_j\in\pi_{a,\delta}$,
of the system described by the equations
\begin{equation} \label{closedloopsampleddelay0001} 
\begin{array}{ll}
\overset{.}{x}\left(t\right)=f(x_t,k(\hat{x}_{t_{j}},h(x_{t_{j}}))),
& \\  \hat{x}\left(t_{j+1}\right)=\hat{x}\left(t_{j}\right)+\left(t_{j+1}-t_j\right)\widehat{f}(\hat{x}_{t_{j}},k(\hat{x}_{t_{j}},h(x_{t_{j}})),h(x_{t_{j}})),
& \\ 
& \\ \hat{x}_{t_{j}}(\theta)=
\begin{cases}
\hat{x}_0(t_j+\theta), & t_j+\theta\leq0,\\ \\
\hat{x}(t_{k})+\dfrac{t_j+\theta-t_k}{t_{k+1}-t_k}(\hat{x}(t_{k+1})-\hat{x}(t_{k})), & t_j+\theta>0, \\
t_k=\underset{l\in N}{\max}\{t_l\in\pi_{a,\delta}\,:\,t_l\leq t_j+\theta\},
\end{cases}
& \\ 
& \\ \theta\in[-\Delta, 0],\,\,\,\,t\in\left[t_j,t_{j+1}\right),\,\,\,\,t_j\in\pi_{a,\delta},\,\,\,\,j=0,1,...,
& \\ x\left(\tau\right)=x_0\left(\tau\right),\,\,\,\,\hat{x}\left(\tau\right)=\hat{x}_0\left(\tau\right),\,\,\,\,\tau\in\left[-\Delta,0\right].
\end{array}
\end{equation}
From (\ref{result0012ok}) it follows that,
\begin{equation} \label{result0001} 
\begin{array}{ll}
\begin{bmatrix}x_t\\
\hat{x}_{t_{j}}
\end{bmatrix}\in \mathcal{C}^{2n}_{E},\,\,\,\,\forall t\in \mathbb{R}^+,\,\,\,\,\forall t_j\in\pi_{a,\delta}, \\
& \\ \begin{bmatrix}x_t\\
\hat{x}_{t_{j}}
\end{bmatrix}\in \mathcal{C}^{2n}_{r},\,\,\,\,\forall t\geq T,\,\,\,\,\forall t_j\in\pi_{a,\delta},\,\,\,\, t_j\geq T.
\end{array}
\end{equation}
The proof of the theorem is complete.

\end{document}